\documentclass[12pt]{article}
\usepackage{amsmath,amssymb,amsthm,amsxtra,overpic,bbm,bm,epsfig,ulem,subfigure}
\usepackage{ulem}
\usepackage{mathrsfs}
\usepackage{graphicx}
\usepackage{epstopdf}
\usepackage{color,ulem}
\usepackage{comment}
\usepackage{float}
\usepackage{cite}
\def\thefootnote{\fnsymbol{footnote}}

\begin{document}

\vspace{0.2cm}

\begin{center}
{\large\bf Correlation of normal neutrino mass ordering with upper octant of
$\theta^{}_{23}$ and third quadrant of $\delta$ via RGE-induced $\mu$-$\tau$ symmetry breaking}
\end{center}

\vspace{0.2cm}

\begin{center}
{\bf Guo-yuan Huang$^{a}$}\footnote{Email: huanggy@ihep.ac.cn},
~ {\bf Zhi-zhong Xing$^{a,b}$\footnote{Email: xingzz@ihep.ac.cn},
~ {\bf Jing-yu Zhu}$^{a}$}\footnote{Email: zhujingyu@ihep.ac.cn} \\
{$^a$Institute of High Energy Physics, and School of Physical
Sciences, \\ University of Chinese Academy of Sciences, Beijing 100049, China \\
$^b$Center for High Energy Physics, Peking University, Beijing
100080, China}
\end{center}

\vspace{1.5cm}

\begin{abstract}
The recent global analysis of three-flavor neutrino oscillation data
indicates that the {\it normal} neutrino mass ordering is favored over the
inverted one at the $3\sigma$ level, and the best-fit values
of the largest neutrino mixing angle $\theta^{}_{23}$ and the Dirac
CP-violating phase $\delta$ are located in the higher octant and the
third quadrant, respectively. We show that all these important issues can be
naturally explained by the $\mu$-$\tau$ reflection
symmetry breaking of massive neutrinos from a superhigh energy scale
down to the electroweak scale due to the one-loop
renormalization-group equations (RGEs) in the minimal supersymmetric standard
model (MSSM). The complete parameter space is explored {\it for the first time}
in both Majorana and Dirac cases, by allowing the smallest neutrino mass $m^{}_1$
and the MSSM parameter $\tan\beta$ to vary in their reasonable regions.
\end{abstract}

\begin{flushleft}
\hspace{0.8cm} PACS number(s): 14.60.Pq, 11.30.Hv, 11.10.Hi.
\end{flushleft}

\def\thefootnote{\arabic{footnote}}
\setcounter{footnote}{0}

\newpage

\section{Introduction}

The striking phenomena of solar, atmospheric, reactor and accelerator
neutrino oscillations have all been observed in the past twenty
years \cite{PDG}, demonstrating that the standard model (SM) of particle
physics is by no means complete and must be extended to explain both the
origin of finite but tiny neutrino masses and the origin of large lepton
flavor mixing effects. Qualitatively, the smallness of neutrino
masses might be attributed to the existence of some heavy degrees of
freedom at a superhigh energy scale --- a popular idea known as the
seesaw mechanism \cite{SS}; and the largeness of neutrino mixing angles and
CP-violating phases might originate from an underlying flavor symmetry
\cite{F1,F2}, which should also manifest itself at a superhigh
energy scale. A combination of the seesaw and flavor symmetry conjectures
turns out to be the most likely phenomenological way of understanding
what is behind the observed spectrum of neutrino masses and the observed
pattern of lepton flavor mixing. In this case the renormalization-group
equations (RGEs) are imperative to bridge the gap between the
(theoretically suggestive) superhigh and (experimentally measurable)
low energy scales \cite{RGE1,RGE2}.

Since the oscillation experiments are insensitive to the Dirac or
Majorana nature of massive neutrinos, one may describe the link between the three
known neutrinos ($\nu^{}_e$, $\nu^{}_\mu$, $\nu^{}_\tau$) and their mass
eigenstates ($\nu^{}_1$, $\nu^{}_2$, $\nu^{}_3$) in terms of a $3\times 3$
unitarity matrix --- the Pontecorvo-Maki-Nakagawa-Sakata (PMNS) matrix \cite{PMNS}
\begin{eqnarray}
V = \left(\begin{matrix}
c^{}_{12} c^{}_{13} & s^{}_{12} c^{}_{13} &
s^{}_{13} e^{-{\rm i} \delta} \cr -s^{}_{12} c^{}_{23} - c^{}_{12}
s^{}_{13} s^{}_{23} e^{{\rm i} \delta} & c^{}_{12} c^{}_{23} -
s^{}_{12} s^{}_{13} s^{}_{23} e^{{\rm i} \delta} & c^{}_{13}
s^{}_{23} \cr s^{}_{12} s^{}_{23} - c^{}_{12} s^{}_{13} c^{}_{23}
e^{{\rm i} \delta} &- c^{}_{12} s^{}_{23} - s^{}_{12} s^{}_{13}
c^{}_{23} e^{{\rm i} \delta} &  c^{}_{13} c^{}_{23} \cr
\end{matrix} \right) \;
\end{eqnarray}
with $c_{ij}^{} \equiv \cos \theta_{ij}^{}$ and
$s_{ij}^{} \equiv \sin \theta_{ij}^{}$ (for $ij = 12, 13, 23$).
A global analysis of currently available data on
neutrino oscillations indicates that the {\it normal} neutrino
mass ordering ($m^{}_1 < m^{}_2 < m^{}_3$) is favored over the
inverted one ($m^{}_3 < m^{}_1 < m^{}_2$) at the
$3\sigma$ level
\footnote{We admit that currently the inverted neutrino mass ordering is still
allowed at the $2 \sigma$ confidence level, but here we give preference
to the normal ordering.}
the best-fit value of the largest neutrino mixing
angle $\theta^{}_{23}$ is slightly larger than $45^\circ$ (i.e., located
in the higher octant), and the best-fit value of the Dirac
phase $\delta$ is somewhat smaller than $270^\circ$ (i.e., located in the
third quadrant) \cite{Lisi,Valle}.
Since $\theta^{}_{23} = 45^\circ$ and $\delta = 270^\circ$
can naturally be derived from the neutrino mass matrix $M^{}_\nu$ constrained by
the $\mu$-$\tau$ reflection symmetry
\footnote{As in the most literature, here the so-called $\mu$-$\tau$ reflection 
symmetry actually means the $\nu^{}_\mu$-$\nu^{}_\tau$ reflection symmetry in
the neutrino sector. One may refer to the discussions above Eq. (2) or Eq. (9)
in section 2, and to Refs. \cite{Grimus:2012hu,Mohapatra:2015gwa}
for building specific models to realize this interesting discrete flavor
symmetry without involving the charged leptons.}
--- a simple flavor symmetry which
assures $M^{}_\nu$ to be invariant under proper charge-conjugation transformations
of the left- and right-handed neutrino fields \cite{F2,HS,HS2},
it is expected to be the {\it minimal}
flavor symmetry responsible for nearly maximal atmospheric neutrino
mixing and potentially maximal CP violation in neutrino oscillations.
If this simple discrete symmetry is realized at a superhigh energy scale,
such as the seesaw scale with $\Lambda^{}_{\mu\tau} \sim 10^{14}$ GeV, it will be
broken at the electroweak scale $\Lambda^{}_{\rm EW} \sim 10^2$ GeV
owing the RGE running effects,
leading to the deviations of $(\theta^{}_{23}, \delta)$
from $(45^\circ, 270^\circ)$. Whether such quantum corrections evolve in the
{\it right} direction to fit the experimental results of $\theta^{}_{23}$ and
$\delta$ at low energies depends on the neutrino mass ordering and the theoretical
framework accommodating the RGEs \cite{Luo,ZhouYL, RGEMixingScheme ,Zhu},
and thus this dependence provides
an elegant possibility of correlating three burning issues in today's neutrino
physics --- the neutrino mass ordering, the octant of $\theta^{}_{23}$ and the
strength of leptonic CP violation.

Given the fact that a credible global analysis of relevant experimental data
often points to the truth in particle physics
\footnote{One of the successful examples of this kind was the global-fit 
``prediction" for an unsuppressed value of $\theta^{}_{13}$ made
in 2008 \cite{Fogli}, which proved to be essentially true after the direct measurement
of $\theta^{}_{13}$ was reported in 2012 \cite{DYB}. This work has recently been recognized by the prestigious Bruno Pontecorvo Prize.},
it is high time for us to take the latest global-fit results of neutrino oscillations
seriously and investigate their implications without involving any details of
model building. In this paper we show that the normal neutrino mass ordering,
the slightly higher octant of $\theta^{}_{23}$ and the possible location of $\delta$ in
the third quadrant can be naturally correlated and explained via RGE-induced
$\mu$-$\tau$ reflection symmetry breaking of massive neutrinos 
--- namely, 
via the one-loop RGE evolution of neutrino masses and flavor mixing parameters
from $\Lambda^{}_{\mu\tau} \sim 10^{14}$ GeV down
to $\Lambda^{}_{\rm EW} \sim 10^2$ GeV in the minimal supersymmetric standard model (MSSM). This kind of correlation will soon be tested by more accurate experimental data.

Although the same topic has been discussed before, our present work
is different from those previous ones at least in the following aspects:
\begin{itemize}
\item      Based on more reliable experimental data, especially the $3\sigma$
indication of the normal neutrino mass ordering, our work is the first one to
numerically explore the almost complete parameter space in both Majorana and
Dirac cases by allowing the smallest neutrino mass $m^{}_1$ and the MSSM parameter
$\tan\beta$ to vary in their reasonable regions. In comparison, the
previous work like Ref.~\cite{RGE2} has only estimated the RGE correction to
$\theta^{}_{23}$ by assuming $\theta^{}_{13} = 0$ --- an approximation which
has now been excluded. It did not consider all the four distinct cases of the Majorana
phases in the $\mu$-$\tau$ reflection symmetry limit either, nor the RGE
correction to $\delta$.

\item      While all the previous studies just assumed some special values of
$m^{}_1$ and $\tan\beta$ to ``illustrate" a possible correlation of the neutrino
mass ordering with the octant of $\theta^{}_{23}$ and the quadrant of $\delta$
based on the RGE-induced $\mu$-$\tau$ reflection symmetry breaking,
our present one has made remarkable progress in ``figuring out" which part of
the parameter space is favored by current neutrino oscillation data and which
part is disfavored or ruled out. The outcome from our in-depth analysis is therefore
more timely, suggestive and useful for model building.

\item      Our statistic analysis shows that currently the best-fit points
of $\theta^{}_{23}$ and $\delta$ \cite{Lisi,Valle} can be explained by
the $\mu$-$\tau$ reflection symmetry breaking induced by the RGE running from
$\Lambda^{}_{\mu\tau}$ down to $\Lambda^{}_{\rm EW}$, but this simple
flavor symmetry itself is in slight tension (at the $1\sigma$ confidence
level) with the low-energy data. Although the best-fit points may shift
when new data become available, the analysis method itself will remain useful.

\item      The upcoming precision measurement of $\theta^{}_{23}$ and an experimental
determination of $\delta$ will test the scenario under consideration and
help locate the correct region in the parameter space for the smallest
neutrino mass $m^{}_1$, the MSSM parameter $\tan\beta$ and even the Majorana phases.
All the previous works were unable to do this job.
\end{itemize}
This paper is organized as follows.
In section 2 we present the main analytical results of the $\mu$-$\tau$ reflection
symmetry breaking caused by the RGE running effects.
In section 3 the parameter space of $\tan\beta$ and $m^{}_1$
is extensively explored and constrained with the help of current experimental data.
Section 4 is devoted to a summary and some concluding remarks.

\section{RGE-induced $\mu$-$\tau$ reflection symmetry breaking}

\subsection{The Majorana case}

Let us assume that the tiny masses of three known neutrinos
originate from a viable seesaw mechanism at a superhigh energy scale
$\Lambda^{}_{\mu\tau} \sim 10^{14}$ GeV. Without loss of generality, we choose the
basis where the mass eigenstates of three charged leptons are identical
with their flavor eigenstates. In this case only the neutrino sector is
responsible for lepton flavor mixing and CP violation. If the effective
Majorana neutrino mass term is invariant under the charge-conjugation
transformations $\nu^{}_{e \rm L} \leftrightarrow \nu^{\rm c}_{e \rm R}$,
$\nu^{}_{\mu \rm L} \leftrightarrow \nu^{\rm c}_{\tau \rm R}$ and
$\nu^{}_{\tau \rm L} \leftrightarrow \nu^{\rm c}_{\mu \rm R}$, the corresponding
mass matrix must take the form
\begin{eqnarray}
M^{}_\nu \equiv \left( \begin{matrix} \langle m\rangle^{}_{ee} &
\langle m\rangle^{}_{e \mu} & \langle m\rangle^{}_{e \tau} \cr
\langle m\rangle^{}_{e \mu} &
\langle m\rangle^{}_{\mu \mu} & \langle m\rangle^{}_{\mu \tau} \cr
\langle m\rangle^{}_{e \tau} &
\langle m\rangle^{}_{\mu \tau} & \langle m\rangle^{}_{\tau \tau}
\end{matrix} \right) \;
\end{eqnarray}
constrained by $\langle m\rangle^{}_{ee} = \langle m\rangle^{*}_{ee}$,
$\langle m\rangle^{}_{e \mu} = \langle m\rangle^{*}_{e \tau}$,
$\langle m\rangle^{}_{\mu \mu} = \langle m\rangle^{*}_{\tau \tau}$ and
$\langle m\rangle^{}_{\mu \tau} = \langle m\rangle^{*}_{\mu \tau}$
\cite{F2}.
One may diagonalize a generic $3\times 3$ Majorana mass matrix
through $U^\dagger M^{}_\nu U^* = {\rm Diag}
\{m^{}_1, m^{}_2, m^{}_3\}$, where the unitary matrix $U$ can be decomposed as
$U = P^{}_l V P^{}_\nu$ with
$P^{}_l = {\rm Diag}\{e^{{\rm i}\phi^{}_e}, e^{{\rm i}\phi^{}_\mu},
e^{{\rm i}\phi^{}_\tau}\}$ and $P^{}_\nu = {\rm Diag}\{e^{{\rm i}\rho},
e^{{\rm i}\sigma}, 1\}$ being the phase matrices, and
$V$ has been shown in Eq. (1). Taking account of the $\mu$-$\tau$
reflection symmetry of $M^{}_\nu$, we immediately arrive at the constraints on
$U$ as follows: $\theta^{}_{23} = 45^\circ$, $\delta = 90^\circ$
or $270^\circ$, $\rho = 0^\circ$ or $90^\circ$, and $\sigma = 0^\circ$
or $90^\circ$ for the four physical parameters, as well as
$\phi_{e}^{} = 90^\circ$ and $\phi_{\mu}^{} + \phi_{\tau}^{} =0^\circ$ for
the three unphysical phases. Since a clear preference for $\sin\delta <0$
has been obtained from the recent global analysis \cite{Lisi},
it is fairly reasonable for us to focus only on
$\delta = 270^\circ$ in the $\mu$-$\tau$ symmetry limit.

In the framework of the MSSM the evolution of $M^{}_\nu$ from $\Lambda^{}_{\rm \mu\tau}$
down to $\Lambda^{}_{\rm EW}$ through the one-loop RGE can be expressed as
\cite{Ellis}
\begin{eqnarray}
M^{}_\nu(\Lambda^{}_{\rm EW}) = I^{2}_0 \left[T^{}_l \cdot M^{}_\nu (\Lambda^{}_{\mu\tau})
\cdot T^{}_l\right] \;
\end{eqnarray}
with $T^{}_l = {\rm Diag}\{I^{}_e, I^{}_\mu, I^{}_\tau\}$, in which
\begin{eqnarray}
I^{}_0 \hspace{-0.2cm} & = & \hspace{-0.2cm}
\exp\left[+\frac{1}{16\pi^2}\int^{\ln\left(\Lambda^{}_{\mu\tau}/
\Lambda^{}_{\rm EW}
\right)}_0 \left(\frac{3}{5}g^2_1(\chi) + 3 g^2_2(\chi) -
3 y^2_t (\chi) \right) {\rm d}\chi \right] \; ,
\nonumber \\
I^{}_\alpha \hspace{-0.2cm} & = & \hspace{-0.2cm}
\exp\left[-\frac{1}{16\pi^2} \int^{\ln\left(\Lambda^{}_{\mu\tau}/\Lambda^{}_{\rm EW}
\right)}_0 y^2_\alpha (\chi) \ {\rm d}\chi \right] \; .
\end{eqnarray}
Here $\chi = \ln \left(\mu/\Lambda^{}_{\mu\tau}\right)$ with $\mu$ being an arbitrary
renormalization scale between $\Lambda^{}_{\rm EW}$ and $\Lambda^{}_{\mu\tau}$,
$g^{}_{1}$ and $g^{}_{2}$ denoting the gauge couplings, $y^{}_t$ and
$y^{}_\alpha$ (for $\alpha = e, \mu, \tau$) standing for the Yukawa coupling
eigenvalues of the top quark and charged leptons, respectively. The smallness
of $y^{}_e$ and $y^{}_\mu$ assures that $I^{}_e \simeq I^{}_\mu \simeq 1$
holds to an excellent degree of accuracy. Note that
\begin{eqnarray}
\Delta^{}_\tau \equiv 1 - I^{}_\tau \simeq \frac{1}{16\pi^2}
\int^{\ln\left(\Lambda^{}_{\mu\tau}/\Lambda^{}_{\rm EW}
\right)}_0 y^2_\tau (\chi) \ {\rm d}\chi \;
\end{eqnarray}
is also a quite small quantity, but it may affect the running behaviors of those
flavor mixing parameters in an appreciable way \cite{Ellis}. To illustrate, Fig. 1
shows the two-dimensional maps of $\Delta_{\tau}^{}$ (left panel) and $I_0^{}$ (right panel) versus the variables $\Lambda^{}_{\mu\tau}$ and $\tan\beta$. One can see that $I_0^{}$ does not change a lot with different settings of $\Lambda_{\mu\tau}^{}$ and $\tan\beta$. In comparison, $\Delta^{}_{\tau}$ can change from $0.001$ to $0.05$. Note that shifting the energy scale is equivalent to altering $\tan\beta$, and the outputs at $\Lambda^{}_{\mu\tau} = 10^{14}$ GeV and $\Lambda^{}_{\mu\tau} = 10^{16}$ GeV are quite similar in magnitude. If we shift the energy scale from $\Lambda^{}_{\mu\tau} = 10^{14}$ GeV to $\Lambda^{}_{\mu\tau} = 10^{9}$ GeV, then $\Delta^{}_{\tau}$ will lie in the range $(0.001,0.03)$ instead of $(0.001,0.05)$. In the following numerical calculations we shall fix
$\Lambda^{}_{\mu\tau} \sim 10^{14}$ GeV as the $\mu$-$\tau$ flavor symmetry scale.
\begin{figure}[t!]
	\begin{center}
		\subfigure{%
			\hspace{-0cm}
			\includegraphics[scale=0.8]{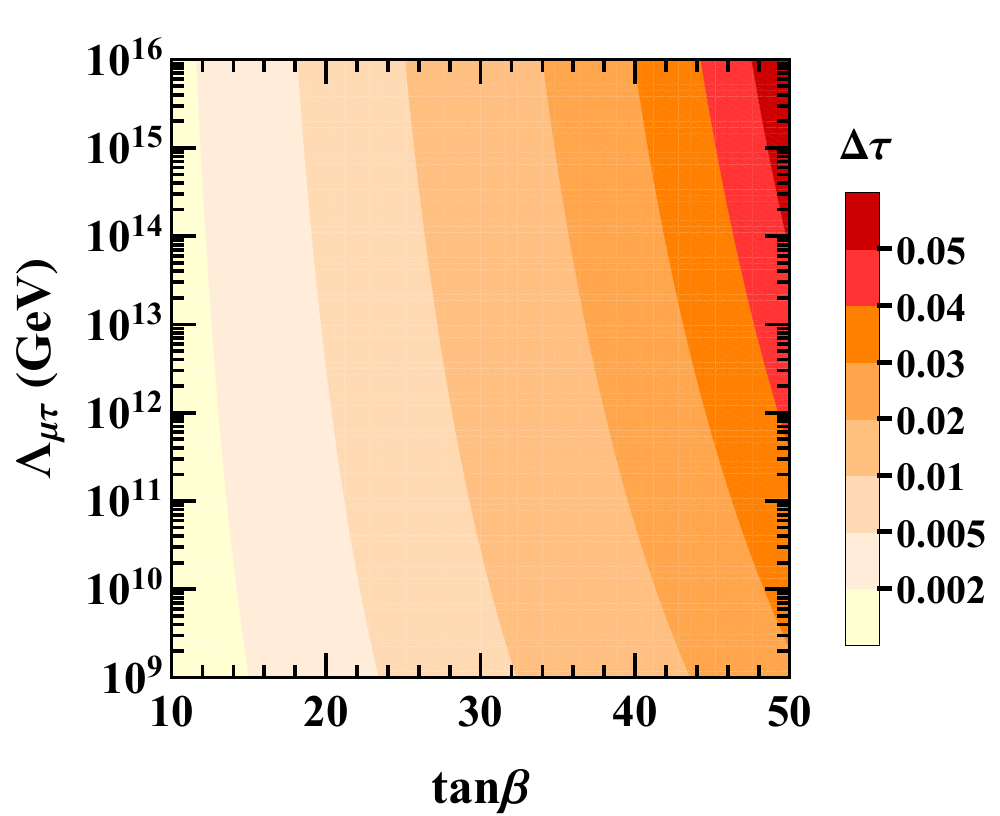}        }%
		\subfigure{%
			\hspace{-0cm}
			\includegraphics[scale=0.78]{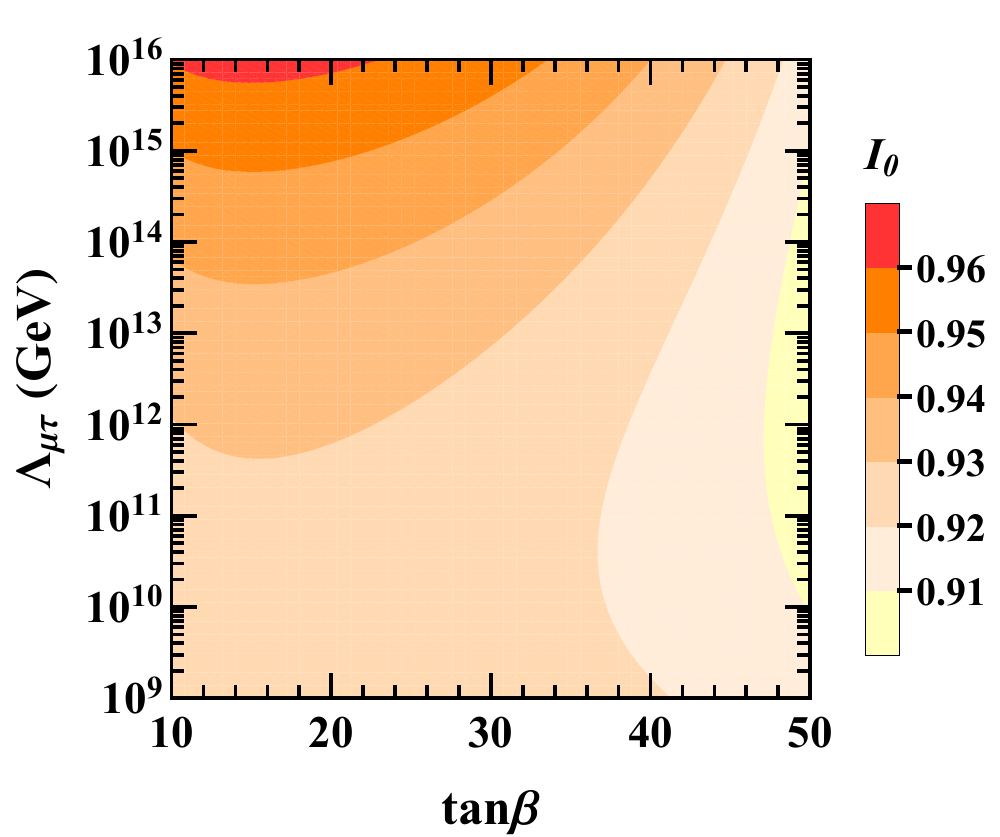}}
	\end{center}
	\vspace{-0.8cm}
	\caption{Possible values of $\Delta_{\tau}^{}$ (left panel) and $I_0^{}$ (right panel) versus the $\mu$-$\tau$ reflection symmetry scale $\Lambda_{\mu\tau}$ and the MSSM parameter $\tan\beta$.}
\end{figure}

One may diagonalize the neutrino mass matrix at $\Lambda^{}_{\rm EW}$ and
then obtain the mass eigenvalues ($m^{}_1, m^{}_2, m^{}_3$),
flavor mixing angles ($\theta^{}_{12}, \theta^{}_{13},
\theta^{}_{23}$) and CP-violating phases ($\delta^{},
\rho^{}, \sigma^{}$). Here we define $\Delta \theta^{}_{ij} \equiv \theta^{}_{ij}
\left(\Lambda_{\rm EW}^{}\right) - \theta^{}_{ij}\left(\Lambda_{\mu\tau}^{}\right)$
(for $ij =12,13,23$), $\Delta \delta \equiv \delta\left(\Lambda_{\rm EW}^{}\right) - \delta\left(\Lambda_{\mu\tau}^{}\right)$,
$\Delta \rho \equiv \rho \left(\Lambda_{\rm EW}^{}\right)
 -\rho\left(\Lambda_{\mu\tau}^{}\right)$ and
$\Delta \sigma \equiv \sigma\left(\Lambda_{\rm EW}^{}\right) -\sigma\left(\Lambda_{\mu\tau}^{}\right)$ to measure the strengths of RGE-induced
corrections to the parameters of $U$. As a good approximation, the three neutrino
masses at $\Lambda^{}_{\rm EW}$ are found to be
\begin{eqnarray}
m_1^{}(\Lambda^{}_{\rm EW})
\hspace{-0.2cm} & \simeq & \hspace{-0.2cm} I^{2}_{0} \left[1 - \Delta^{}_{\tau}
\left(1 - c_{12}^{2} c_{13}^2 \right)\right] m_1^{}(\Lambda_{\mu\tau}^{})  \; ,
\nonumber \\
m_2^{}(\Lambda^{}_{\rm EW})
\hspace{-0.2cm} & \simeq & \hspace{-0.2cm} I^{2}_{0}  \left[1 - \Delta^{}_{\tau}
\left(1-s^{2}_{12} c^{2}_{13}\right)\right] m_2^{}(\Lambda_{\mu\tau}^{}) \; ,
\nonumber \\
m_3^{}(\Lambda^{}_{\rm EW})
\hspace{-0.2cm} & \simeq & \hspace{-0.2cm} I^{2}_{0}  \left[1 -\Delta^{}_{\tau}
c^{2}_{13}\right] m_3^{}(\Lambda_{\mu\tau}^{}) \; ,
\end{eqnarray}
in which $\theta^{}_{12}$ and $\theta^{}_{13}$ take their values at $\Lambda^{}_{\rm EW}$.
Unless otherwise specified, the nine physical flavor parameters appearing in the
subsequent text and equations are all the ones at $\Lambda^{}_{\rm EW}$. In a
reasonable analytical approximation we can also arrive at
\footnote{Note that our analytical results are not exactly the same as those obtained in
Ref. \cite{ZhouYL}, where a different phase convention for the PMNS matrix has
been used.}
\begin{eqnarray}
\Delta \theta_{12}^{} \hspace{-0.2cm} & \simeq &
\hspace{-0.2cm} \frac{\Delta_{\tau}^{}}{2} c_{12}^{}
s_{12}^{} \left[s_{13}^2\left(\zeta_{31}^{\eta_{\rho}^{}}
- \zeta_{32}^{\eta_{\sigma}^{}} \right)
+ c_{13}^2 \zeta_{21}^{-\eta_{\rho}^{} \eta_{\sigma}^{}} \right] \;,
\nonumber\\
\Delta \theta_{13}^{} \hspace{-0.2cm} & \simeq &
\hspace{-0.2cm} \frac{\Delta_{\tau}^{}}{2} c_{13}^{}
s_{13}^{} \left(c_{12}^2 \zeta_{31}^{\eta_{\rho}^{}}
+ s_{12}^2 \zeta_{32}^{\eta_{\sigma}^{}}\right) \;,
\nonumber\\
\Delta \theta_{23}^{} \hspace{-0.2cm} & \simeq &
\hspace{-0.2cm} \frac{\Delta_{\tau}^{}}{2} \left(s_{12}^2
\zeta_{31}^{-\eta_{\rho}^{}} + c_{12}^2
\zeta_{32}^{-\eta_{\sigma}^{}}\right) \;
\end{eqnarray}
for the deviations of three flavor mixing angles between $\Lambda^{}_{\rm EW}$ and
$\Lambda^{}_{\mu\tau}$; and
\begin{eqnarray}
\Delta \delta \hspace{-0.2cm} & \simeq &
\hspace{-0.2cm}  \frac{\Delta_{\tau}^{}}{2}
\left[\frac{c_{12}^{} s_{12}^{}}{s_{13}^{}}
\left(\zeta_{32}^{-\eta_{\sigma}^{}} -
\zeta_{31}^{-\eta_{\rho}^{}}\right) -
\frac{s_{13}^{}}{c_{12}^{} s_{12}^{}}
\left(c_{12}^4 \zeta_{32}^{-\eta_{\sigma}} -
s_{12}^4 \zeta_{31}^{-\eta_{\rho}^{}} +
\zeta_{21}^{\eta_{\rho}^{} \eta_{\sigma}}\right)\right] \;,
\nonumber\\
\Delta \rho \hspace{-0.2cm} & \simeq &
\hspace{-0.2cm}  \Delta_{\tau}^{}
\frac{c_{12}^{} s_{13}^{}}{s_{12}^{}} \left[ s_{12}^{2}
\left(\zeta_{31}^{-\eta_{\rho}} -
\zeta_{32}^{-\eta_{\sigma}^{}}\right) +\frac{1}{2}
\left(\zeta_{32}^{-\eta_{\sigma}^{}} +
\zeta_{21}^{\eta_{\rho}^{} \eta_{\sigma}}\right)\right] \;,
\nonumber \\
\Delta \sigma \hspace{-0.2cm} & \simeq &
\hspace{-0.2cm}  \Delta_{\tau}^{}
\frac{s_{12}^{} s_{13}^{}}{2 c_{12}^{}}
\left[ s_{12}^2 \left(\zeta_{21}^{\eta_{\rho}^{}\eta^{}_\sigma}
- \zeta_{31}^{-\eta_{\rho}^{}}\right) -
c_{12}^{2} \left(2 \zeta_{32}^{-\eta_{\sigma}^{}} -
\zeta_{31}^{-\eta_{\rho}^{}} - \zeta_{21}^{
	\eta_{\rho}^{} \eta_{\sigma}^{}}\right)\right] \;
\end{eqnarray}
for the deviations of three CP-violating phases between $\Lambda^{}_{\rm EW}$ and
$\Lambda^{}_{\mu\tau}$,
where $\eta_{\rho}^{} \equiv \cos 2\rho = \pm 1$ and $\eta_{\sigma}^{} \equiv
\cos 2\sigma = \pm 1$ denote the possible options of $\rho$ and $\sigma$ in
their $\mu$-$\tau$ symmetry limit at $\Lambda^{}_{\mu\tau}$,
and the ratios $\zeta^{}_{ij} \equiv (m^{}_i - m^{}_j)/
(m^{}_i + m^{}_j)$ are defined with $m^{}_i$ and $m^{}_j$ at $\Lambda^{}_{\rm EW}$
(for $i, j = 1, 2, 3$). In obtaining Eqs. (6)---(8) the $\mu$-$\tau$ reflection
symmetry conditions $\theta^{}_{23}(\Lambda_{\mu\tau}^{}) = 45^\circ$ and $\delta(\Lambda_{\mu\tau}^{}) = 270^\circ$ have been applied too.

\subsection{The Dirac case}

Although the Majorana nature of massive neutrinos is well motivated
from a theoretical point of view, it is also interesting to consider
the possibility of a pure Dirac mass term for three known neutrinos and combine
it with a certain flavor symmetry which can be realized at a superhigh
energy scale $\Lambda^{}_{\mu\tau}$ \cite{Zhu}. In this case the $\mu$-$\tau$ reflection
symmetry means that the Dirac neutrino mass term is invariant under the
charge-conjugation transformations
$\nu^{}_{e \rm L} \leftrightarrow (\nu^{}_{e \rm L})^{\rm c}$,
$\nu^{}_{\mu \rm L} \leftrightarrow (\nu^{}_{\tau \rm L})^{\rm c}$ and
$\nu^{}_{\tau \rm L} \leftrightarrow (\nu^{}_{\mu \rm L})^{\rm c}$ for
the left-handed neutrino fields and
$N^{}_{e \rm R} \leftrightarrow (N^{}_{e \rm R})^{\rm c}$,
$N^{}_{\mu \rm R} \leftrightarrow (N^{}_{\tau \rm R})^{\rm c}$ and
$N^{}_{\tau \rm R} \leftrightarrow (N^{}_{\mu \rm R})^{\rm c}$ for the
right-handed neutrino fields. The resultant Dirac neutrino mass matrix is
\begin{eqnarray}
M^{}_\nu \equiv \left( \begin{matrix} \langle m\rangle^{}_{ee} &
\langle m\rangle^{}_{e \mu} & \langle m\rangle^{}_{e \tau} \cr
\langle m\rangle^{}_{\mu e} &
\langle m\rangle^{}_{\mu \mu} & \langle m\rangle^{}_{\mu \tau} \cr
\langle m\rangle^{}_{\tau e} &
\langle m\rangle^{}_{\tau \mu} & \langle m\rangle^{}_{\tau \tau}
\end{matrix} \right) \;
\end{eqnarray}
which is constrained by $\langle m\rangle^{}_{ee} = \langle m\rangle^{\ast}_{ee}$,
$\langle m\rangle^{}_{e\mu} = \langle m\rangle^{\ast}_{e\tau}$,
$\langle m\rangle^{}_{\mu e} = \langle m\rangle^{\ast}_{\tau e}$,
$\langle m\rangle^{}_{\mu\tau} = \langle m\rangle^{\ast}_{\tau\mu}$ and
$\langle m\rangle^{}_{\mu\mu} = \langle m\rangle^{\ast}_{\tau\tau}$
\cite{Zhu}. One can see that $M^{}_\nu$ is in general neither symmetric
nor Hermitian, even though it possesses the $\mu$-$\tau$ reflection
symmetry. But one may diagonalize the Hermitian combination $M^{}_\nu M^\dagger_\nu$
by means of the unitary transformation
$U^\dagger M^{}_\nu M^\dagger_\nu U = {\rm Diag}\{m^2_1, m^2_2, m^2_3\}$,
where $U = P^{}_l V$ with $P^{}_l = {\rm Diag}\{e^{{\rm i}\phi^{}_e},
e^{{\rm i}\phi^{}_\mu}, e^{{\rm i}\phi^{}_\tau}\}$ being an unphysical
phase matrix and $V$ taking the form given in Eq. (1). Then the
$\mu$-$\tau$ reflection symmetry constraints on $M^{}_\nu$ naturally lead us to
the constraints on the PMNS matrix $U$; namely, $\theta^{}_{23} = 45^\circ$,
$\delta = 90^\circ$ or $270^\circ$, and $2\phi^{}_e - \phi^{}_\mu - \phi^{}_\tau
= 180^\circ$ at $\Lambda_{\mu\tau}^{}$. Given the global-fit preference for $\sin\delta <0$ \cite{Lisi},
we focus only on the possibility of $\delta (\Lambda_{\mu\tau}^{}) = 270^\circ$.

In the MSSM the evolution of the Dirac neutrino mass matrix
$M^{}_\nu$ from $\Lambda^{}_{\mu\tau}$ down to $\Lambda^{}_{\rm EW}$ via
the one-loop RGE can be described as \cite{Zhu}
\begin{eqnarray}
M^{}_\nu (\Lambda^{}_{\rm EW}) = I^{}_0 \left[T^{}_l \cdot M^{}_\nu (\Lambda^{}_{\mu\tau})
\right] \; ,
\end{eqnarray}
where the definitions of $I^{}_0$ and $T^{}_l$ are the same as those in Eqs. (3)
and (4). Diagonalizing $M^{}_\nu (\Lambda^{}_{\rm EW})$ will yield
the seven physical flavor parameters at $\Lambda^{}_{\rm EW}$. By use of
the same notations as in the Majorana case, let us summarize our approximate
analytical results:
\begin{eqnarray}
m^{}_1(\Lambda^{}_{\rm EW}) \hspace{-0.2cm} & \simeq & \hspace{-0.2cm}
I^{}_{0} \left[1 - \frac{1}{2} \Delta^{}_{\tau}
\left(1 - c^{2}_{12} c^{2}_{13} \right)\right] m^{}_1(\Lambda_{\mu\tau}^{}) \; ,
\nonumber \\
m^{}_2(\Lambda^{}_{\rm EW}) \hspace{-0.2cm} & \simeq & \hspace{-0.2cm}
I^{}_{0} \left[1 - \frac{1}{2}\Delta^{}_{\tau}
\left(1 - s^{2}_{12} c^{2}_{13}\right)\right] m_2^{}(\Lambda_{\mu\tau}^{}) \; ,
\nonumber \\
m^{}_3(\Lambda^{}_{\rm EW}) \hspace{-0.2cm} & \simeq & \hspace{-0.2cm}
I^{}_{0} \left[1 - \frac{1}{2}\Delta^{}_{\tau}
c^{2}_{13}\right] m_3^{}(\Lambda_{\mu\tau}^{}) \;
\end{eqnarray}
for the masses of three neutrinos; and
\begin{eqnarray}
\Delta\theta^{}_{12} \hspace{-0.2cm} & \simeq & \hspace{-0.2cm}
\frac{\Delta^{}_{\tau}}{4} s^{}_{12}c^{}_{12}
\left[ c^{2}_{13}\left(\zeta_{21}^{} + \zeta_{21}^{-1}\right) -
s^{2}_{13} \left(\zeta_{32}^{} + \zeta_{32}^{-1} -\zeta_{31}^{} -
\zeta_{31}^{-1} \right) \right] \; ,
\nonumber \\
\Delta\theta^{}_{13} \hspace{-0.2cm} & \simeq & \hspace{-0.2cm}
\frac{\Delta^{}_{\tau}}{4} s^{}_{13}c^{}_{13}
\left[ s^{2}_{12} \left(\zeta_{32}^{} + \zeta_{32}^{-1}\right) +
c^{2}_{12}\left(\zeta_{31}^{} + \zeta_{31}^{-1}\right)
\right] \; ,
\nonumber \\
\Delta\theta^{}_{23} \hspace{-0.2cm} & \simeq & \hspace{-0.2cm}
\frac{\Delta^{}_{\tau}}{4} \left[ c^2_{12}\left(\zeta_{32}^{} +
\zeta_{32}^{-1}\right) +s^2_{12} \left(\zeta_{31}^{} +
\zeta_{31}^{-1}\right) \right] \;
\end{eqnarray}
for the deviations of three flavor mixing angles between $\Lambda^{}_{\rm EW}$ and
$\Lambda^{}_{\mu\tau}$; and
\begin{eqnarray}
\Delta\delta^{} \hspace{-0.2cm} & \simeq & \hspace{-0.2cm}
\frac{\Delta^{}_{\tau}}{4} \left[
\frac{c_{12}^{} \left(s^2_{12}
-c^2_{12} s^2_{13}\right)}{s_{12}^{} s^{}_{13}}
\left(\zeta_{32}^{} + \zeta_{32}^{-1}\right) -
\frac{s_{12}^{} \left(c^2_{12}
-s^2_{12} s^2_{13}\right)}{c_{12}^{} s^{}_{13}}
\left(\zeta_{31}^{} + \zeta_{31}^{-1}\right) \right.
\nonumber \\
& & \hspace{-0.2cm} \left.
-\frac{s_{13}^{}}{c_{12}^{} s_{12}^{}} \left(\zeta_{21}^{} +
\zeta_{21}^{-1}\right)\right] \;
\end{eqnarray}
for the deviation of $\delta$ between $\Lambda^{}_{\rm EW}$ and
$\Lambda^{}_{\mu\tau}$. In obtaining Eqs. (11)---(13) we have taken
into account the $\mu$-$\tau$ reflection symmetry conditions
$\theta^{}_{23}(\Lambda_{\mu\tau}^{}) = 45^\circ$ and
$\delta(\Lambda_{\mu\tau}^{}) = 270^\circ$.

The analytical approximations made in Eqs. (7) and (8)
for Majorana neutrinos or those made in Eqs. (12) and (13)
for Dirac neutrinos are instructive and helpful for understanding
the RGE corrections to relevant flavor mixing and CP-violating
parameters, but the accuracy will be quite poor if the neutrino
masses are strongly degenerate. In the subsequent section we shall
numerically evaluate the effects of $\mu$-$\tau$ reflection symmetry
breaking and explore the allowed parameter space to fit current
experimental data.

\section{Numerical exploration of the parameter space}

In the framework of the MSSM we numerically run the RGEs from $\Lambda_{\mu\tau}^{} \sim 10^{14}~{\rm GeV}$ down to $\Lambda^{}_{\rm EW} \sim 10^{2}~{\rm GeV}$ by taking account of the initial conditions
$\theta^{}_{23}=45^{\circ}$ and $\delta = 270^{\circ}$ as well as the initial values of
$\rho$ and $\sigma$ in four different cases ---
{\bf Case A}: $\rho=\sigma=0^\circ$; {\bf Case B}: $\rho=\sigma=90^\circ$; {\bf Case C}: $\rho=0^{\circ}$ and $\sigma=90^\circ$; {\bf Case D}: $\rho=90^{\circ}$ and $\sigma=0^\circ$. For any given values of the MSSM parameter $\tan\beta$ and the smallest neutrino mass $m^{}_1$ at $\Lambda^{}_{\rm EW}$, the other relevant neutrino oscillation parameters like $\{ \sin^2{\theta_{12}}, ~\sin^2{\theta_{13}}, ~\Delta m_{\rm sol}^2, ~\Delta m_{\rm atm}^2\}$ at $\Lambda^{}_{\mu\tau}$  are scanned over properly wide ranges with the help of the MultiNest program \cite{Feroz}. Here we have adopted the notations $\Delta m_{\rm sol}^2 \equiv m_2^2-m_1^2$ and $\Delta m_{\rm atm}^2 \equiv m_3^2- (m_1^2+m_2^2)/2$ as defined in Ref.~\cite{Lisi}
\footnote{In the normal neutrino mass ordering case one may therefore express $m^{}_2$
and $m^{}_3$ in terms of $m^{}_1$ as follows:
$m^{}_2 = \sqrt{m^2_1 + \Delta m^2_{\rm sol}}~$ and
$m^{}_3 = \sqrt{m^2_1 + 0.5 \Delta m^2_{\rm sol} + \Delta m^2_{\rm atm}}~$.}.
For each scan, the neutrino flavor parameters at $\Lambda^{}_{\rm EW}$ are yielded and
they are immediately compared with their global-fit values by minimizing
\begin{align}
\chi^2 \equiv \sum_i \frac{\left(\xi_i - \overline{\xi}_i\right)^2}{\sigma_i^2} \; ,
\end{align}
where $\xi^{}_i$'s stand for the parameters at $\Lambda^{}_{\rm EW}$ produced from the RGE evolution, $\overline{\xi}^{}_i$'s represent the best-fit values of the global analysis, and $\sigma^{}_i$'s are the corresponding symmetrized $1\sigma$ errors (i.e., $\sigma^{}_i = (\sigma_i^{+} + \sigma_i^{-})/2$).
In our numerical calculations we concentrate on the normal neutrino mass ordering
\footnote{It is not only our phenomenological preference but also a recent $3\sigma$ global-fit indication that the true neutrino mass spectrum should exhibit a
normal ordering $m^{}_1 < m^{}_2 < m^{}_3$.}
and adopt the best-fit values and the $1\sigma$-level deviations of
the six neutrino oscillation parameters \cite{Lisi}:
\begin{eqnarray}
&& \sin^2\theta^{}_{12} = 3.04^{+0.14}_{-0.13} \times 10^{-1} \; ,
\nonumber \\
&& \sin^2\theta^{}_{13} = 2.14^{+0.09}_{-0.07}\times 10^{-2} \; ,
\nonumber \\
&& \sin^2\theta^{}_{23}
= 5.51^{+0.19}_{-0.70} \times 10^{-1} \; ,
\nonumber \\
&& \delta = 1.32^{+0.23}_{-0.18} \times \pi \; ,
\nonumber \\
&& \Delta m^2_{\rm sol} =
7.34^{+0.17}_{-0.14} \times 10^{-5} ~{\rm eV}^2 \; ,
\nonumber \\
&& \Delta m^2_{\rm atm} = 2.455^{+0.035}_{-0.032} \times 10^{-3} ~{\rm eV}^2 \; . \hspace{0.6cm}
\end{eqnarray}
In this case the smallest neutrino mass $m^{}_1$ is allowed to take values in the range $[0, 0.1]~{\rm eV}$, and the MSSM parameter $\tan{\beta}$ may vary from $10$ to $50$ based on a reasonable phenomenological argument
\footnote{Note that $\tan{\beta} > 50$ is disfavored because the heavy-quark Yukawa couplings
would fall into the non-perturbative region, while for $\tan{\beta} < 10$ the RGE-induced corrections to the relevant neutrino parameters are negligibly small and thus less interesting for our purpose.}.
It should be pointed out that our numerical results are independent of the analytical
approximations made in the last section, but the latter will be helpful for understanding
some salient features of the former.

\subsection{The Majorana case}

The strategy of our numerical analysis is rather straightforward.
Let us first examine how significantly $\theta^{}_{23}$ and $\delta$ at $\Lambda^{}_{\rm EW}$ can deviate from their initial values at $\Lambda^{}_{\mu\tau}$, incorporating with the recent global-fit results.
To do so we only need to take into account the global-fit information about the parameter
set $\xi = \{ \sin^2{\theta_{12}}, ~\sin^2{\theta_{13}}, ~\Delta m_{\rm sol}^2, ~\Delta m_{\rm atm}^2\}$. One will see that the RGE running effects always push $\theta^{}_{23}$ to the higher octant and in most cases lead $\delta$ to the third quadrant --- just the right direction as indicated by the best-fit
values of these two quantities \cite{Lisi}. After this preliminary diagnosis is made,
the experimental information on $\theta^{}_{23}$ and $\delta$ will be included so as to evaluate the statistical compatibility between the RGE-triggered $\mu$-$\tau$ reflection symmetry breaking and the global-fit values of $\theta^{}_{23}$ and $\delta$.
    \begin{figure}[t]
    \begin{center}
     \hspace{-0.1cm}
    \includegraphics[width=0.5\textwidth]{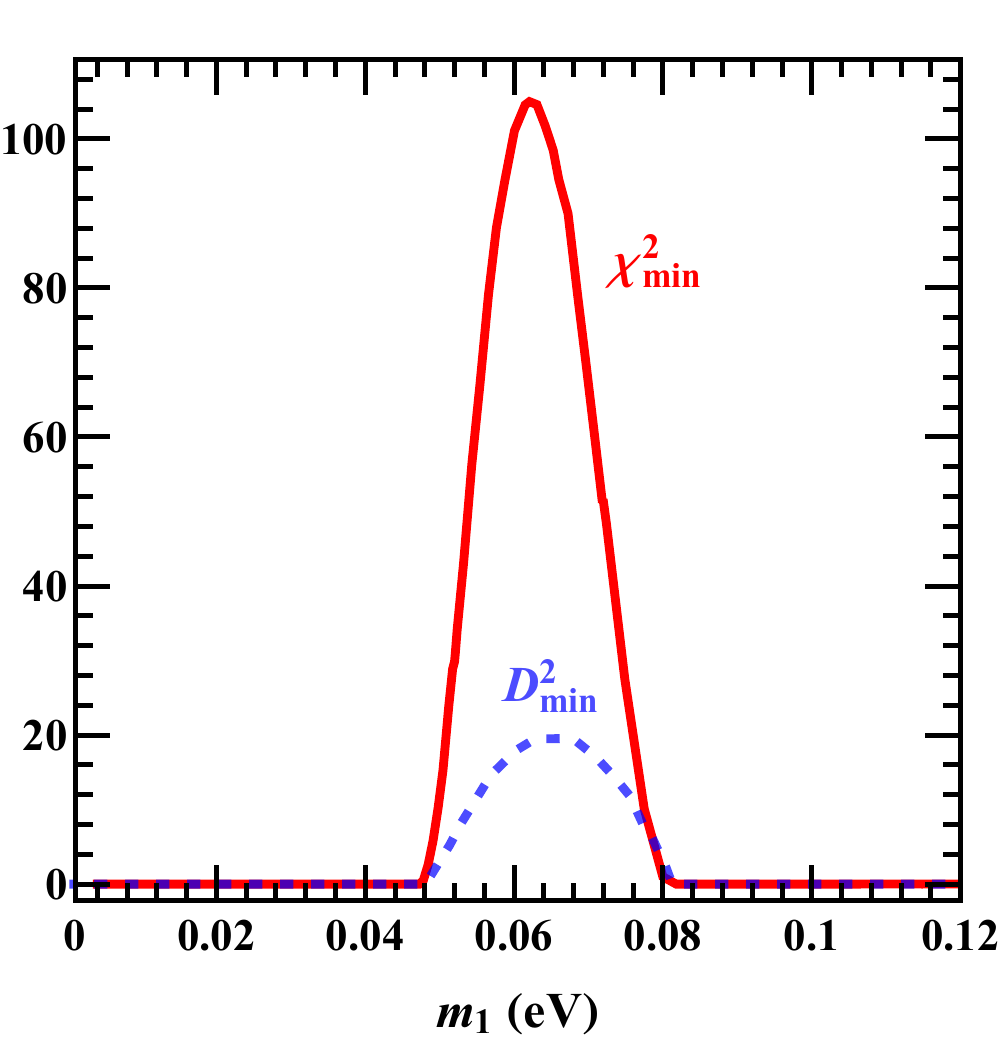}
    \end{center}
    \vspace{-0.8cm}
     \caption{The behavior of $\chi^2_{\rm min}$ (or $D^2_{\rm min}$) with respect to the neutrino mass $m^{}_1$ for {\bf Case C} (i.e., $\rho = 0^\circ$ and $\sigma = 90^\circ$ at
     $\Lambda^{}_{\mu\tau}$), where $\tan\beta = 50$ has been typically input.}
    \label{fig:chi2}
    \end{figure}

For each given value of $m^{}_1$ or $\tan{\beta}$ at $\Lambda_{\rm EW}^{}$, we obtain the associated $\chi^2_{\rm min}$ from Eq. (14) which is minimized over the chosen parameter set $\xi$. In Fig.~\ref{fig:chi2} we show $\chi^2_{\rm min}$ with respect to $m^{}_1$
for {\bf Case C} as an example. It is obvious that $\chi^2_{\rm min}$ can reach 0 (i.e., the best-fit point) in most cases, but for $m^{}_1 \in [0.05, 0.08]~{\rm eV}$ the value of $\chi^2_{\rm min}$ bumps up to nearly $100$. $\chi^2_{\rm min} > 0$ means that the $\mu$-$\tau$ reflection symmetry limit at $\Lambda^{}_{\mu\tau}$ cannot be touched if one runs the RGEs inversely --- starting from the best-fit points of six neutrino oscillation parameters at $\Lambda^{}_{\rm EW}$. This observation will also be true even if one allows $\theta^{}_{23}$ and $\delta$ to take arbitrary values at $\Lambda^{}_{\rm EW}$. The reason should be ascribed to the nontrivial differential structures of the RGEs \cite{RGE1,RGE2}, especially when the evolution becomes wild in a narrow parameter space. In such a case the Majorana phases $\rho$ and $\sigma$ play a potentially significant role. We have demonstrated that $\chi^2_{\rm min}$ in the bumped region is actually the local minimum. The reliability of this result is also verified by running the system from $\Lambda^{}_{\rm EW}$ up to $\Lambda^{}_{\mu\tau}$. The squared distance to the $\mu$-$\tau$ reflection symmetry limit $\overline{\eta} \equiv \{\theta^{}_{23}=45^{\circ}, \delta = 270^{\circ}, \rho = 0^\circ, \sigma = 90^{\circ}\}$ at $\Lambda_{\mu\tau}^{}$ is defined as
\begin{align}
D^2 \equiv \sum_i \frac{\left(\eta^{}_i - \overline{\eta}^{}_i\right)^2}{1000} \; ,
\end{align}
where the relevant parameters $\eta^{}_{i}$'s are obtained by running the
RGEs from $\Lambda^{}_{\rm EW}$ up to $\Lambda_{\mu\tau}^{}$ with $\{
\sin^2{\theta_{12}}, ~\sin^2{\theta_{13}}, ~\Delta m_{\rm sol}^2, ~\Delta
m_{\rm atm}^2\}$ all taking their best-fit values at $\Lambda_{\rm EW}^{}$.
As shown in Fig.~\ref{fig:chi2}, $D^2_{\rm min}$ has a similar bump whose
range coincides with that of $\chi^2_{\rm min}$.
In Eq. (16) the normalization factor $1/1000$ is introduced
to simply make the amplitude of the $D^2$ bump the same order of magnitude 
as that of $\chi^2_{\rm min}$. This check confirms the
observation we have obtained by running the RGEs from
$\Lambda^{}_{\mu\tau}$ down to $\Lambda^{}_{\rm EW}$.
    \begin{figure}[t]
    \begin{center}
     \hspace{-0.1cm}
    \includegraphics[width=1\textwidth]{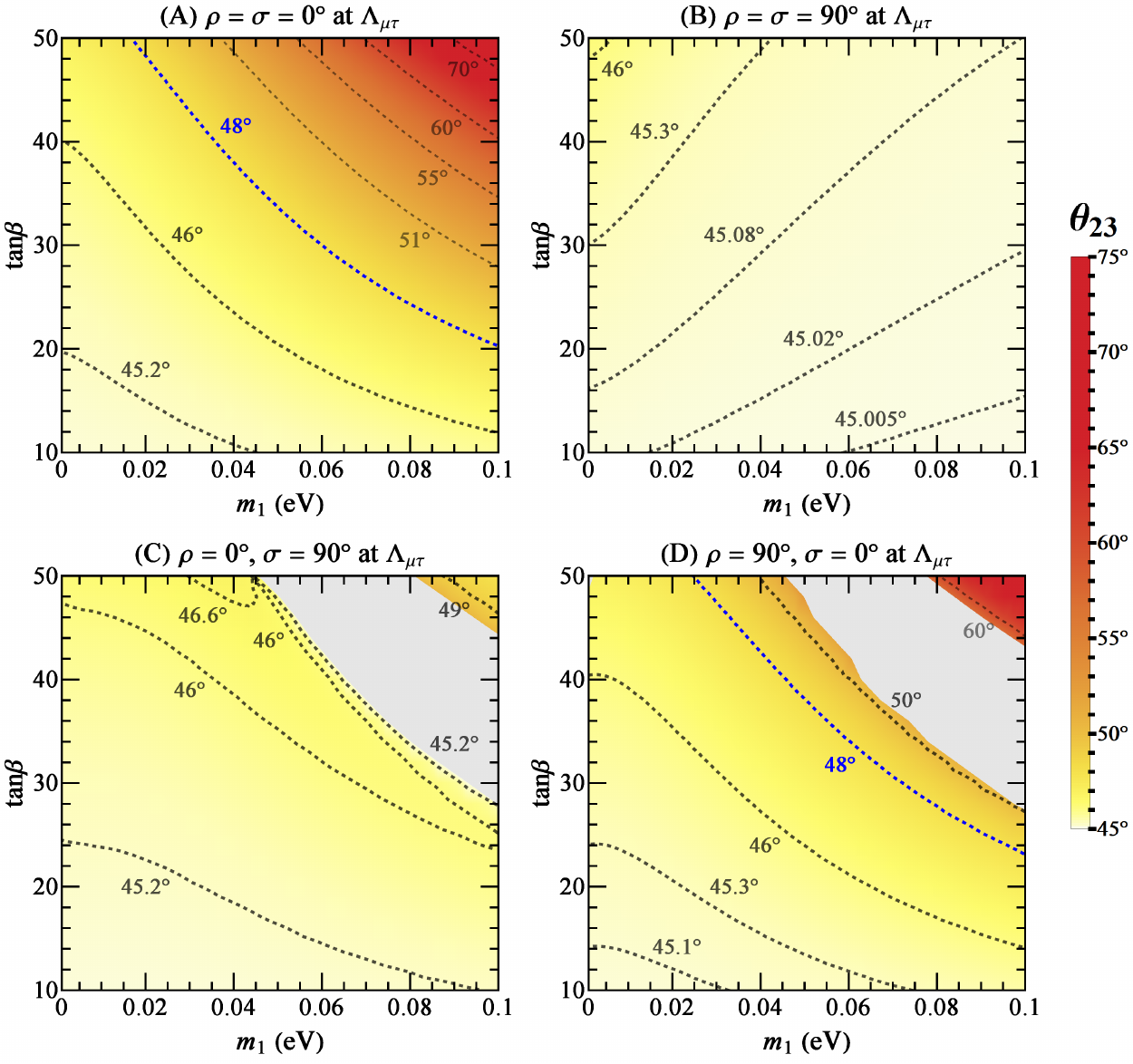}
    \end{center}
    \vspace{-0.8cm}
    \caption{The allowed region of $\theta^{}_{23}$ at $\Lambda^{}_{\rm EW}$ due to the RGE-induced $\mu$-$\tau$ reflection symmetry breaking, where the dashed curves are the contours for some typical values of $\theta^{}_{23}$ and the blue one is compatible with the best-fit result of $\theta^{}_{23}$ obtained in Ref. \cite{Lisi}.}
    \label{fig:mt23}
    \end{figure}
 	\begin{figure}[t!]
    \begin{center}
     \hspace{-0.1cm}
    \includegraphics[width=1\textwidth]{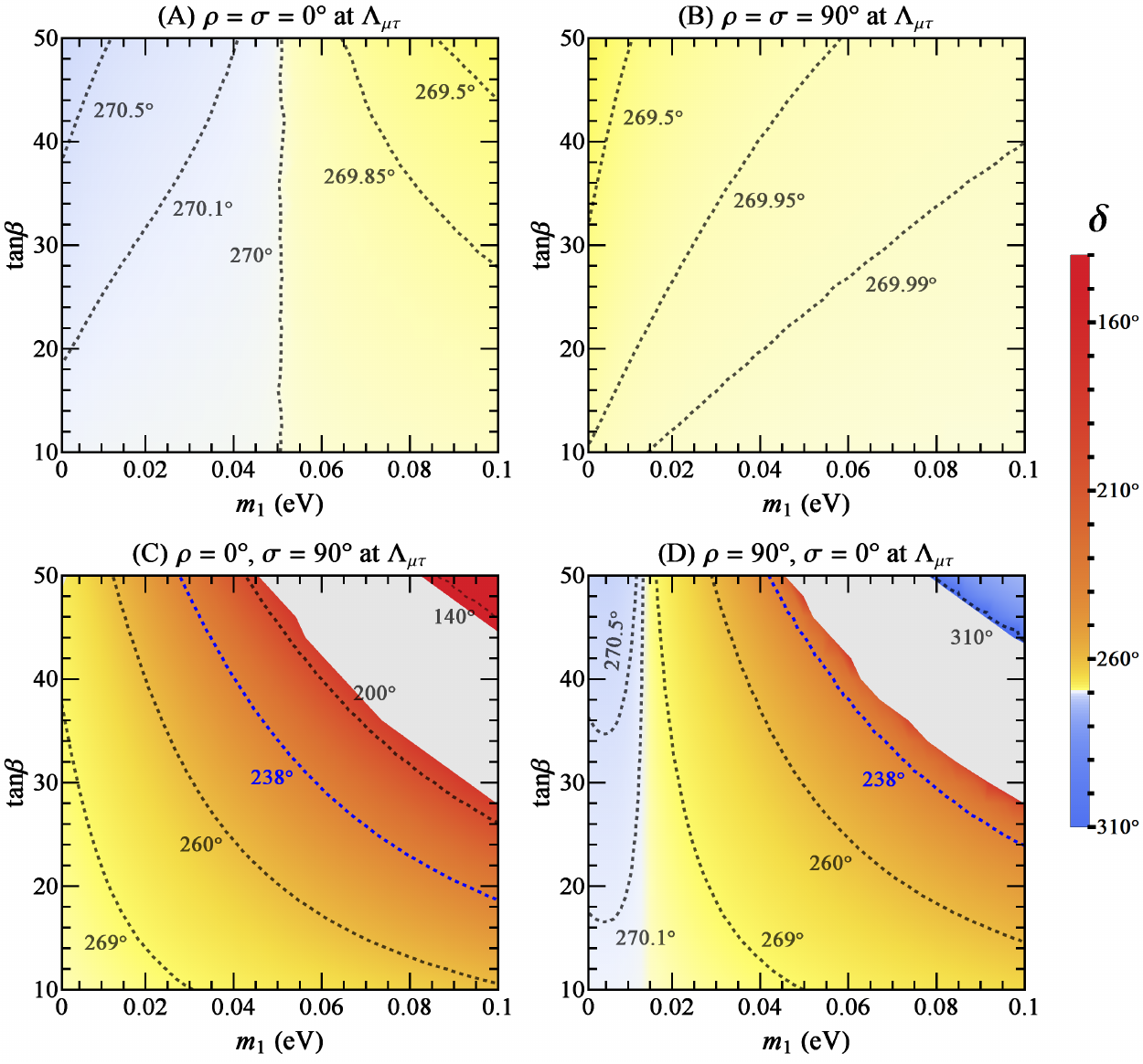}
    \end{center}
    \vspace{-0.8cm}
   \caption{The allowed region of $\delta$ at $\Lambda^{}_{\rm EW}$ due to the RGE-induced $\mu$-$\tau$ reflection symmetry breaking, where the dashed curves are the contours for some typical values of $\delta$ and the blue one is compatible with the best-fit result of $\delta$ obtained in Ref. \cite{Lisi}.}
    \label{fig:md}
    \end{figure}
    \begin{figure}[t!]
    \begin{center}
     \hspace{-0.1cm}
    \includegraphics[width=1\textwidth]{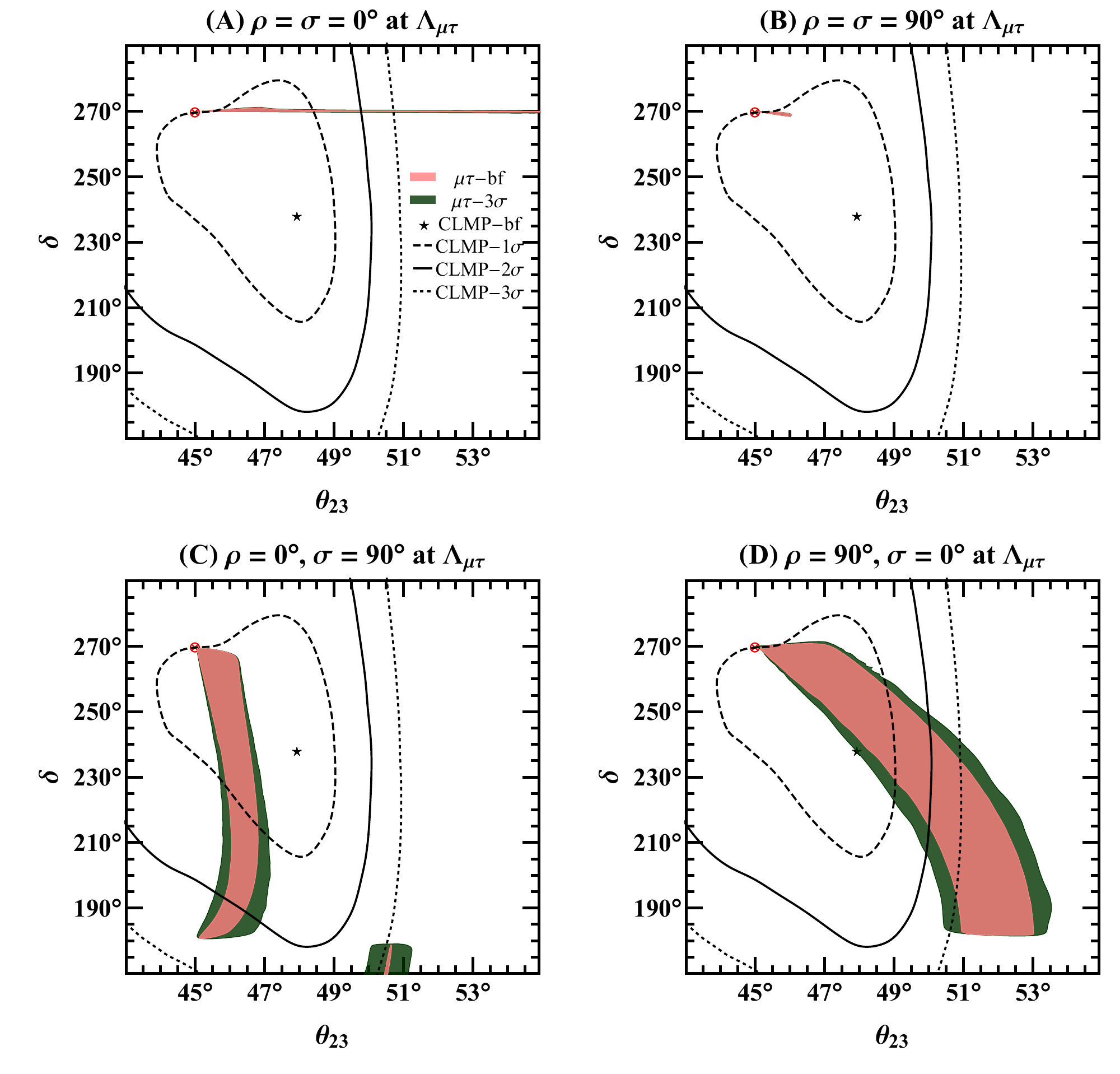}
    \end{center}
    \vspace{-0.8cm}
    \caption{The correlation between $\theta^{}_{23}$ and $\delta$ at
    $\Lambda_{\rm EW}^{}$ as compared with the recent global-fit
    results (abbreviated as ``CLMP") \cite{Lisi}, where $m^{}_1$ and
   	$\tan{\beta}$ have been marginalized respectively over $[0,
   	0.1]~{\rm eV}$ and $[10, 50]$, and the red circled cross
    {\footnotesize $\boldsymbol\otimes$} 	
   	stands for the point $(\theta^{}_{23},\delta) =
   	(45^{\circ},270^{\circ})$. The pink region is allowed for
   	$\theta^{}_{23}$ and $\delta$ when $\{ \sin^2{\theta^{}_{12}},
   	~\sin^2{\theta^{}_{13}}, ~\Delta m_{\rm sol}^2, ~\Delta m_{\rm
   	atm}^2\}$ at $\Lambda_{\rm EW}$ take their best-fit values, and
   	the green region is allowed when these four observables  deviate from
   	their best-fit values by $3\sigma$ level (i.e., $\chi^2 = 11.83$ for
   	two degrees of freedom).}
    \label{fig:lisi}
    \end{figure}

Fig.~\ref{fig:mt23} shows the RGE-corrected result of $\theta^{}_{23}$ at $\Lambda^{}_{\rm EW}$ for different values of $m^{}_1$ and $\tan{\beta}$ with $\chi^2 = 0$, where the four possible options of initial $(\rho,\sigma)$ at $\Lambda_{\mu\tau}^{}$ have been considered. Note that the boundary conditions for the RGEs include both the initial values of $\{\theta^{}_{23}, \delta, \rho, \sigma\}$ at $\Lambda^{}_{\mu\tau}$ and the experimental constraints on $\{ \sin^2{\theta^{}_{12}}, ~\sin^2{\theta^{}_{13}}, ~\Delta m_{\rm sol}^2, ~\Delta m_{\rm atm}^2\}$ at $\Lambda^{}_{\rm EW}$, which are all specified in our numerical calculations. In this case one should keep in mind that the low- and high-scale boundary requirements are likely to be so strong that the RGEs do not have a realistic solution --- the gray-gap regions in Fig.~\ref{fig:mt23} ({\bf Cases C} and {\bf D}), corresponding to the $\chi^2_{\rm min} > 0$ bump region in Fig.~\ref{fig:chi2}. Some more comments on Fig.~\ref{fig:mt23} are in order.
\begin{itemize}
\item The gray-gap regions in {\bf Cases C} and {\bf D} are due to the $\chi^2_{\rm min}$-bump as shown in Fig.~\ref{fig:chi2}, but there is not such a gap for {\bf Cases A} and {\bf B}. In {\bf Case A} the RGE running effect is quite significant, and $\theta^{}_{23}$ may run to almost $75^{\circ}$ for $m^{}_1 \simeq 0.1~{\rm eV}$ and $\tan\beta \simeq 50$. Of course, such evolution will be strongly constrained by the experimental information on $\theta^{}_{23}$ which has not yet been included into our analysis. In contrast, $\theta^{}_{23}$ is not sensitive to the RGE corrections in {\bf Case B}, and it maximally changes only about $1^{\circ}$. As for {\bf Cases C} and {\bf D}, if one conservatively requires $m^{}_1 < 0.07~{\rm eV}$ from the present cosmological bound \cite{Ade:2015xua}, it will be possible for $\theta^{}_{23}$ to run to $46.6^{\circ}$ and $50^{\circ}$, respectively. The best-fit value $\theta^{}_{23} \simeq48^{\circ}$ \cite{Lisi} can be easily reached in {\bf Cases A} and {\bf D}.

\item The RGE correction to $\theta^{}_{23}$ illustrated in Fig.~\ref{fig:mt23} can be well understood with the help of the analytical approximations made in Eq. (7), only if the neutrino masses are not so degenerate and the RGE evolution is not so strong. Eq. (7) tells us that the sign of $\Delta \theta^{}_{23}$ is positive, because $\Delta^{}_{\tau}$ and $\zeta^{}_{31} \simeq \zeta^{}_{32}$ are all positive for the normal neutrino mass ordering. The factor $\Delta^{}_{\tau}$ is essentially proportional to $\tan^2\beta$ as a result of $y^2_{\tau} \propto (1 + \tan^2\beta) \simeq \tan^2\beta$ for $\tan\beta \gtrsim 10$, and therefore $\Delta \theta^{}_{23}$ always increases with $\tan^2\beta$. On the other hand, the dependence of $\theta^{}_{23}$ on the neutrino mass $m^{}_1$ is different in the four options of $\rho$ and $\sigma$ at $\Lambda^{}_{\mu\tau}$. For example, $\zeta^{-\eta_{\rho}}_{31}$ is proportional to $m^{}_1$ for $\rho (\Lambda^{}_{\mu\tau}) = 0^{\circ}$, but it is proportional to $1/m^{}_1$ when $\rho (\Lambda^{}_{\mu\tau}) = 90^{\circ}$. In the region of small $m^{}_1$ and $\tan\beta$, the radiative correction to $\theta^{}_{23}$ is proportional to $m^{}_1$ for {\bf Cases A}, {\bf C} and {\bf D}, but it is inversely proportional to $m^{}_1$ in {\bf Case B} with $\eta^{}_{\rho} = \eta^{}_{\sigma} = -1$.
\end{itemize}
       \begin{figure}[t]
    \begin{center}
     \hspace{-0.1cm}
    \includegraphics[width=0.5\textwidth]{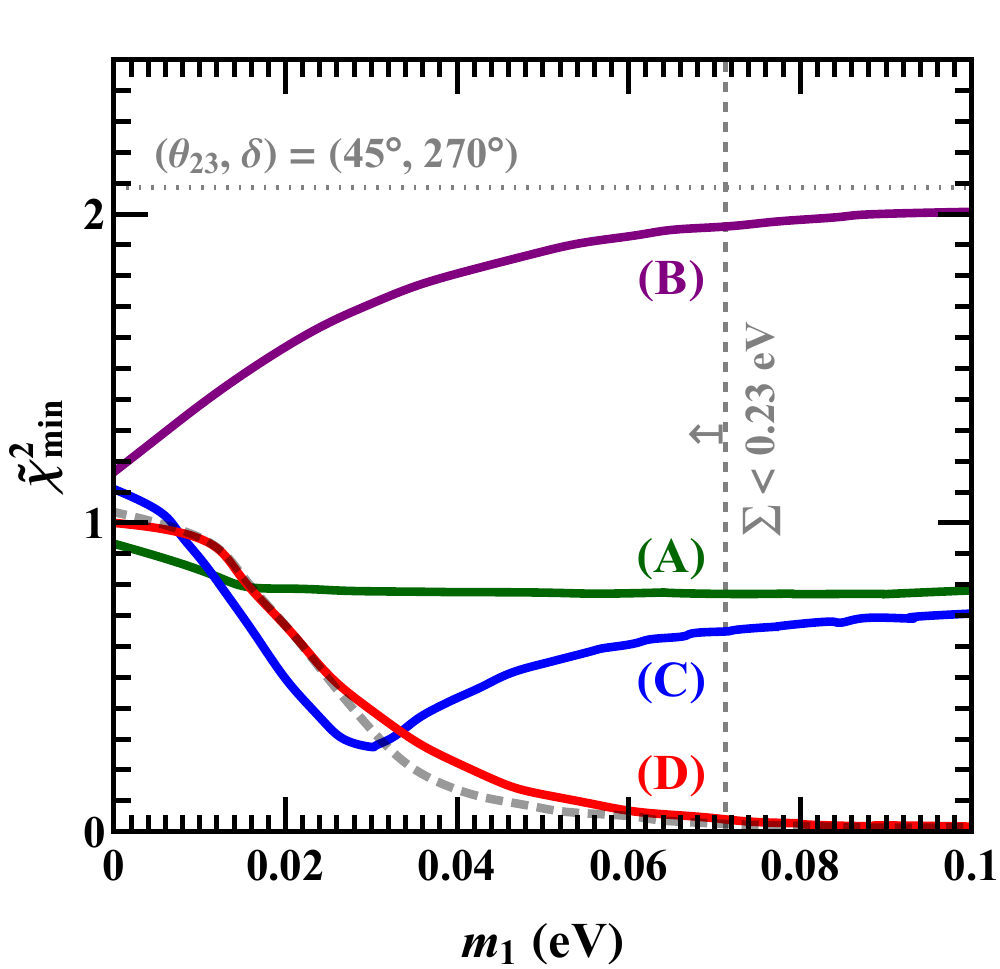}
    \end{center}
    \vspace{-0.8cm}
    \caption{The minimal $\tilde{\chi}^2$ by marginalizing over $\tan\beta \in [10,50]$. The colored curves stand for the four Majorana cases, and the dashed gray curve denotes the Dirac case. The vertical dashed line is derived from the cosmological limit on the sum of the neutrino masses \cite{Ade:2015xua}, and the horizontal line represents $\tilde{\chi}^2_{\rm min}$ of the point $(\theta^{}_{23},\delta) = (45^{\circ}, 270^{\circ})$ at $\Lambda_{\rm EW}^{}$.}
    \label{fig:chimin}
    \end{figure}

In Fig.~\ref{fig:md} we plot the allowed region of $\delta$ at $\Lambda_{\rm EW}^{}$. Note that for each point in the $m^{}_1$-$\tan\beta$ plane, $\delta$ and $\theta^{}_{23}$ are determined at the same time. Some discussions are in order.
\begin{itemize}
\item The RGE-induced corrections to $\delta$ in {\bf Cases A} and {\bf B} are very weak, only about $0.5^{\circ}$. Even though the higher octant of $\theta^{}_{23}$ (including its best-fit value) can be easily reached in {\bf Case A}, it is impossible to approach the best-fit value of $\delta$ (i.e., $\delta \simeq 238^{\circ}$). But the best-fit value of $\delta$ can be reached in both {\bf Case C} and {\bf Case D}. It deserves to highlight {\bf Case D} in which the best-fit point $(\theta^{}_{23},\delta) \simeq (48^{\circ}, 238^{\circ})$ is reachable from the same settings of $m^{}_{1}$ and $\tan\beta$.

\item Similar to the case of $\theta^{}_{23}$, the radiative correction to $\delta$ is also proportional to $\tan^2\beta$ regardless of its sign. But the dependence of $\delta$ on $m^{}_1$ in Eq. (8) is not so straightforward as that of $\theta^{}_{23}$ in Eq. (7). There are two terms in $\Delta\delta$, one is enhanced by $1/\sin\theta^{}_{13}$ and the other is suppressed by $\sin\theta^{}_{13}$, but the latter can become dominant in some cases. In {\bf Case A} the first term $\propto 1/\sin\theta^{}_{13}$ is positive and dominant when the neutrino mass $m^{}_1$ is relatively small, while the second term $\propto \sin\theta^{}_{13}$ is negative and will gradually dominate when the value of $m^{}_1$ increases. These analytical features can explain the numerical evolution behavior of $\delta$ for {\bf Case A} shown in Fig.~\ref{fig:md}. In {\bf Case B} with $\eta^{}_{\rho}=\eta^{}_{\sigma} = -1$, both of the two terms of $\Delta\delta$ are negative and inversely proportional to $m^{}_1$. Note that the first term of $\Delta\delta$ in either {\bf Case A} or {\bf Case B} is suppressed due to the cancellation between $\zeta^{-\eta^{}_\sigma}_{32}$ and $\zeta^{-\eta^{}_\rho}_{31}$, and this largely explains the smallness of $\Delta\delta$. In {\bf Case C} (or {\bf Case D}) the first term of $\Delta\delta$ is negative (or positive) and initially dominant, but the second term containing $\zeta_{21}^{\eta_{\rho} \eta_{\sigma}} = \zeta_{21}^{-1}$ will eventually dominate for relatively bigger values of $m^{}_1$. Hence the RGE-induced corrections to $\delta$ in these two cases can be greatly enhanced by $1/\Delta m^{2}_{\rm sol}$.
\end{itemize}

To see the correlation between $\theta^{}_{23}$ and $\delta$ at $\Lambda^{}_{\rm EW}$, let us marginalize $m^{}_1$ and $\tan\beta$ over the reasonable ranges $m^{}_1 \in [0,0.1]~{\rm eV}$ and $\tan{\beta} \in [10,50]$. Our numerical outputs are summarized in Fig.~\ref{fig:lisi},
where the recent global-fit results \cite{Lisi} are plotted as black lines
with the $1\sigma$ (dashed), $2\sigma$ (solid) and
$3\sigma$ (dotted) contours. The corresponding
best-fit point of their analysis is marked as the black star. We notice that the $\mu$-$\tau$ reflection symmetry point $(\theta^{}_{23},\delta) = (45^{\circ},270^{\circ})$ at $\Lambda^{}_{\rm EW}$, which is marked as the red 
circled cross in the plot, is on the dashed contour. The latter means that $\theta^{}_{23}(\Lambda^{}_{\rm EW}) = 45^\circ$ and $\delta (\Lambda^{}_{\rm EW}) = 270^{\circ}$ are statistically disfavored at the $1\sigma$ level \cite{Lisi}. The $\theta^{}_{23}$-$\delta$ correlation at low energies, which arises from RGE-induced $\mu$-$\tau$ symmetry breaking,
is described by the pink or green region. In the pink region the best-fit values of $\{ \sin^2{\theta^{}_{12}}, ~\sin^2{\theta^{}_{13}}, ~\Delta m_{\rm sol}^2, ~\Delta m_{\rm atm}^2\}$ can be simultaneously reached (i.e., $\chi^2_{\rm min} = 0$).
If the value of $\chi^2_{\rm min}$ is relaxed to $11.83$ (i.e., the $3\sigma$ confidence level for two degrees of freedom), the wider green region of $\theta^{}_{23}$ and $\delta$ will be allowed. In the two upper panels of Fig.~\ref{fig:lisi} which correspond to {\bf Cases A} and {\bf B}, the allowed range of $\delta$ is very narrow --- this feature is compatible with the two upper panels of Fig.~\ref{fig:md} where $\delta$ varies less than $1^{\circ}$. In these two cases the green region almost overlaps the pink region. In the two lower panels of Fig.~\ref{fig:lisi} corresponding to {\bf Cases C} and {\bf D}, the RGE-induced corrections are significant. Note that there is a separate shaded region around $\theta^{}_{23} \simeq 50^{\circ}$ in Fig.~\ref{fig:lisi}(C), and it is associated with the small upper-right corner of the parameter space in Fig.~\ref{fig:mt23}(C) or Fig.~\ref{fig:md}(C). There is a similar separate shaded region in {\bf Case D}, but it is outside the chosen ranges of $\theta^{}_{23}$ and $\delta$ in plotting Fig.~\ref{fig:lisi}(D) and its confidence level is much weaker --- outside the $3\sigma$ region of the global analysis.
    \begin{figure}[t]
    \begin{center}
    \subfigure{%
     \hspace{-0.1cm}
    \includegraphics[width=0.48\textwidth]{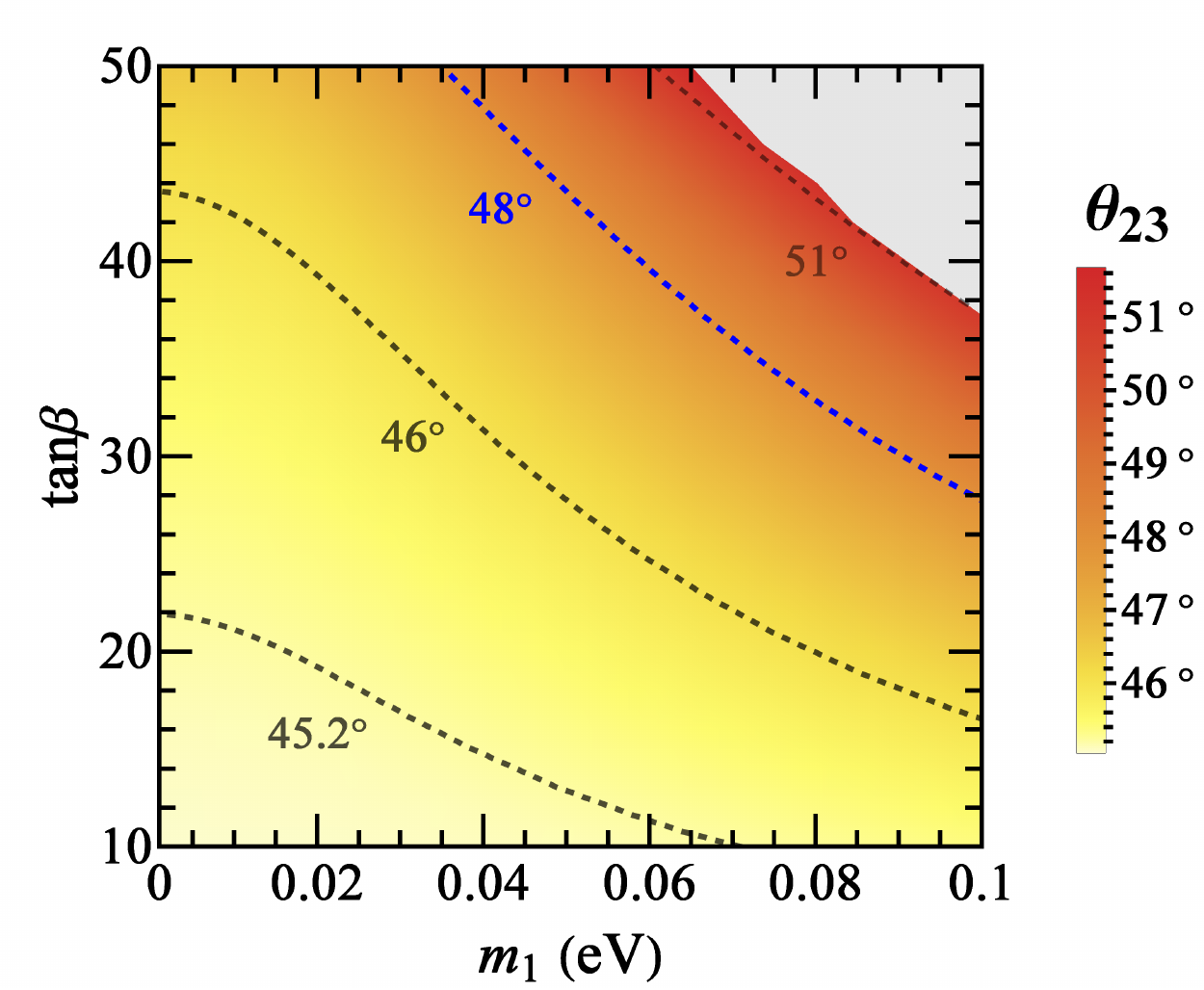}        }%
    \subfigure{%
    \hspace{-0.1cm}
    \includegraphics[width=0.49\textwidth]{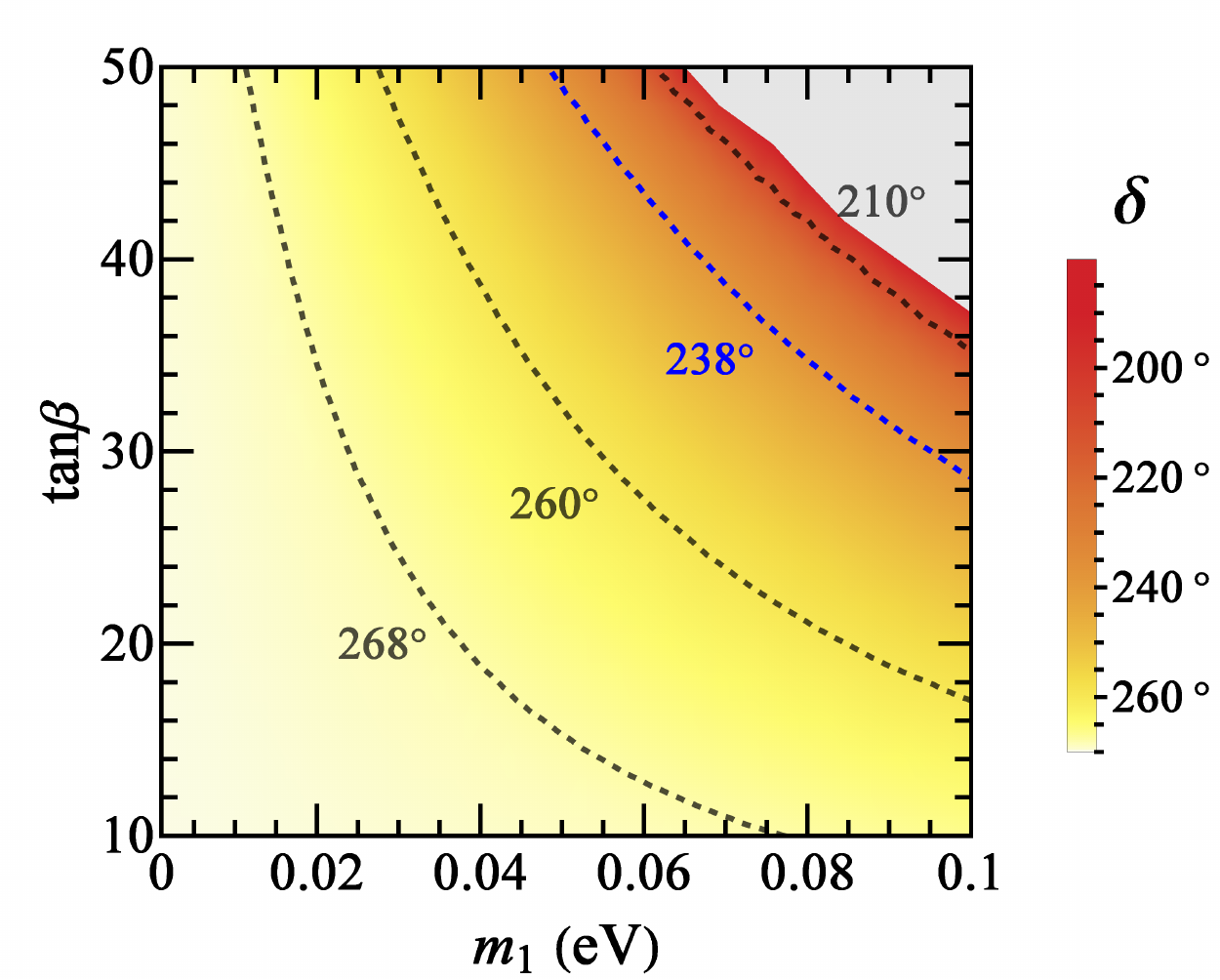}        }
    \end{center}
    \vspace{-0.8cm}
    \caption{In the Dirac case the allowed regions of $\theta^{}_{23}$ (left panel) and $\delta$ (right panel) at $\Lambda^{}_{\rm EW}$ due to the RGE-induced $\mu$-$\tau$ reflection symmetry breaking, where the dashed curves are the contours for some typical values of $\theta^{}_{23}$ and $\delta$, and the blue one is compatible with the best-fit result of $\theta^{}_{23}$ or $\delta$  obtained in Ref. \cite{Lisi}.}
    \label{fig:diracnh}
    \end{figure}

To numerically judge the compatibility between our $\mu$-$\tau$ symmetry
breaking scenario and current experimental data, or how well the RGE-triggered
$\mu$-$\tau$ reflection symmetry breaking effect can fit the complete set of data at
$\Lambda_{\rm EW}^{}$, now we include the global-fit information about
$\theta^{}_{23}$ and $\delta$. Namely, we define $\tilde{\chi}^2 \equiv \chi^2 +
\chi^2_{\theta^{}_{23}} + \chi^2_{\delta}$, in which $\chi^2$ is
formed with the parameter set $\{ \sin^2{\theta^{}_{12}}, ~\sin^2{\theta^{}_{13}},
~\Delta m_{\rm sol}^2, ~\Delta m_{\rm atm}^2\}$ just as before,
and $\chi^2_{\theta^{}_{23}}$ and $\chi^2_{\delta}$ are the
contributions from $\theta^{}_{23}$ and $\delta$ respectively. In this case
the minimum of $\tilde{\chi}^2$ can be calculated for each of the four
cases of $\rho$ and $\sigma$ at $\Lambda^{}_{\mu\tau}$ by marginalizing
the relevant quantities over $\tan\beta \in [10,50]$, and the result is
plotted in Fig.~\ref{fig:chimin} as a function of $m^{}_1$.
For the special point $(\theta_{23},\delta) =
(45^{\circ},270^{\circ})$ at $\Lambda_{\rm EW}^{}$, the corresponding
$\tilde{\chi}^2_{\rm min}$ reads 2.08. Fig.~\ref{fig:chimin} shows
the very good reduction of $\tilde{\chi}^2_{\rm min}$ by incorporating the
RGE running effect in the framework of the MSSM.
Among the four cases under discussion, the red curve for {\bf Case D}
with $\rho = 90^{\circ}$ and $\sigma = 0^{\circ}$ at $\Lambda_{\mu\tau}^{}$ is most
outstanding. Even given the Planck limit on the sum of the neutrino
masses $\sum \equiv m_1^{} + m_2^{} + m_3^{}< 0.23~{\rm eV}$ at the
$95\%$ confidence level \cite{Ade:2015xua}, it is still
possible to reduce the value of $\tilde{\chi}^2_{\rm min}$ 
to almost $0.05$.
    \begin{figure}[t]
    \begin{center}
     \hspace{-0.1cm}
    \includegraphics[width=0.5\textwidth]{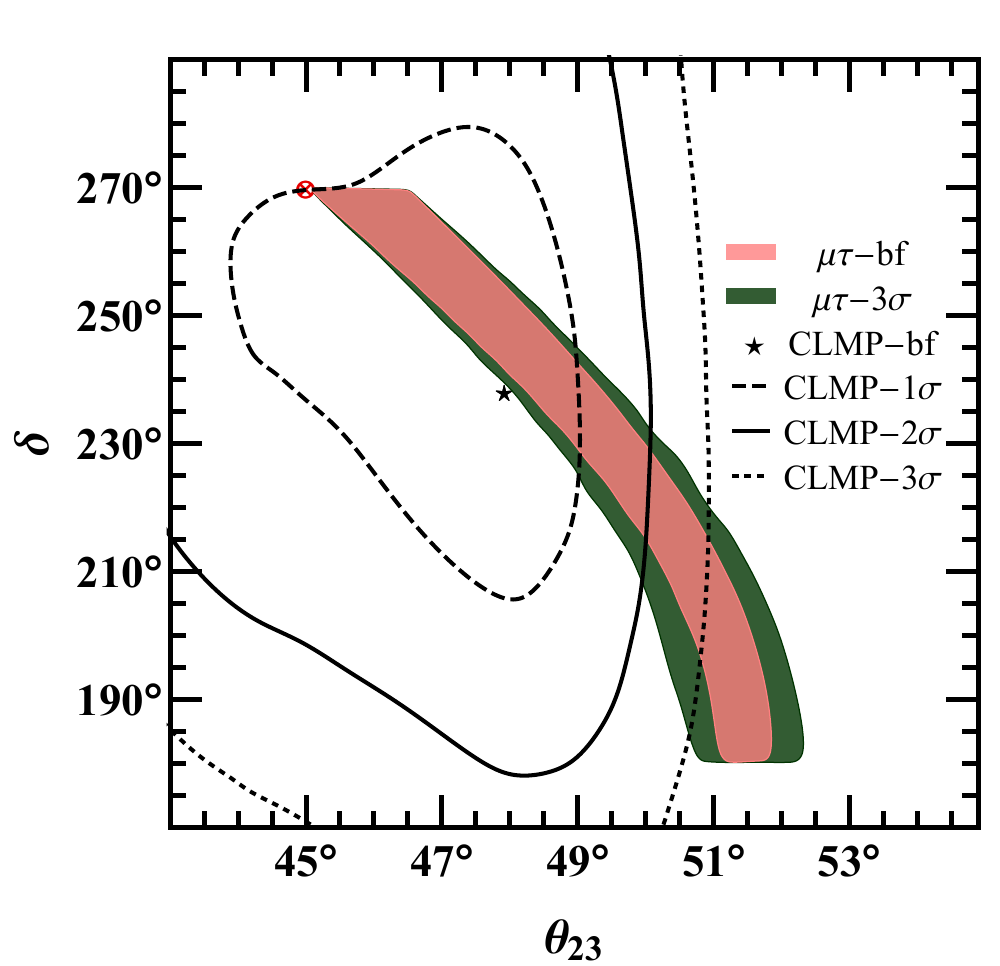}
    \end{center}
    \vspace{-0.8cm}
    \caption{The correlation of the broken values of $(\theta_{23},\delta)$ for the Dirac case. The notations stay the same with Fig.~\ref{fig:lisi}.}
    \label{fig:diracnh2}
    \end{figure}

\subsection{The Dirac case}

Since there is only a single CP-violating phase in the Dirac case, it is much easier to
do a numerical analysis of the parameter space which is constrained by both the RGE-induced
$\mu$-$\tau$ reflection symmetry breaking effect and the recent global fit of neutrino
oscillation data. Fig.~\ref{fig:diracnh} shows the allowed regions of $\theta^{}_{23}$ and $\delta$ at $\Lambda_{\rm EW}$, and their intimate correlation is illustrated in Fig.~\ref{fig:diracnh2}. Here the evolution behaviors of these two parameters can be understood in a way more straightforward than in the Majorana case, simply because of the absence of the two Majorana phases. In the leading-order approximation the analytical expressions of $\Delta\theta^{}_{23}$ and $\Delta\delta$ in Eqs. (12) and (13) are simplified to
\begin{eqnarray}
&& \Delta\theta^{}_{23} \simeq
\frac{\Delta^{}_{\tau}}{2} \frac{m_2^2+m_3^2}{\Delta m^2_{\rm atm}} \; ,
\nonumber \\
&& \Delta\delta^{} \simeq
-\frac{\Delta^{}_{\tau}}{2} \frac{s_{13}^{}}{c_{12}^{} s_{12}^{}}
\frac{m_1^2+m_2^2}{\Delta m^2_{\rm sol}} \; . \hspace{0.6cm}
\end{eqnarray}
It becomes obvious that $\left(\theta_{23}^{} , \delta\right)$ may have larger deviations from $\left(45^{\circ}, 270^{\circ}\right)$ for
bigger values of $m^{}_1$ and $\tan\beta$. In particular, $\theta^{}_{23}$ and $\delta$
are always located in the upper octant and the third quadrant,
respectively. The existence of the gray regions in
Fig.~\ref{fig:diracnh} is for the same reason as that in the Majorana case, as we have
discussed above. Similar to {\bf Case D} in the Majorana scenario, the RGE-induced
corrections can take $(\theta^{} _{23},\delta)$ very close to their best-fit point, and
the corresponding $\tilde{\chi}^2_{\rm min}$ is shown in Fig.~\ref{fig:chimin} as the
gray dashed line.

\section{Concluding remarks}

In neutrino physics it is usually necessary (and popular) to introduce
some heavy degrees of freedom and certain flavor symmetries at a superhigh
energy scale, so as to explain the tiny masses of three known neutrinos
and the striking pattern of lepton flavor mixing observed at low energies.
In this case it is also necessary to make use of the RGEs as a powerful tool
to bridge the gap between these two considerably different energy scales.
Such RGE-induced quantum corrections may naturally break the given flavor
symmetry and thus lead to some phenomenologically interesting consequences,
including a possible correlation between the neutrino mass ordering and
flavor mixing parameters.

In this work we have considered the intriguing $\mu$-$\tau$ reflection symmetry and
its RGE-induced breaking as an instructive playground
to realize the aforementioned idea, especially in view of the fact that
the recent global analysis of neutrino oscillation data has
indicated a remarkable preference for the normal neutrino mass ordering
at the $3\sigma$ level together with a slightly higher octant of
$\theta^{}_{23}$ and the possible location of $\delta$ in the third quadrant.
We have shown that all these important issues can be naturally correlated and explained
by the RGE-triggered $\mu$-$\tau$ reflection symmetry breaking from a
superhigh energy scale $\Lambda^{}_{\mu\tau} \sim 10^{14}$ GeV down to
the electroweak scale $\Lambda^{}_{\rm EW} \sim 10^2$ GeV in the MSSM.
Different from those previous attempts in this connection,
our study represents the first numerical exploration of the
complete parameter space in both Majorana and Dirac cases by allowing
the smallest neutrino mass $m^{}_1$ and the MSSM parameter $\tan\beta$ to
respectively vary in their reasonable regions $[0,0.1]$ eV
and $[10,50]$. We believe that direct measurements of the neutrino
mass ordering and precision measurements of $\theta^{}_{23}$ and $\delta$
in the near future will test the simple but suggestive scenario under consideration.

Of course, some of our main observations are subject to the
MSSM itself and current best-fit values of $\theta^{}_{23}$ and $\delta$.
The reason why we have chosen the MSSM instead of the SM is three-fold: (a)
in the SM it is extremely difficult to generate an appreciable value of
$\Delta\theta^{}_{23}$ via the RGE-induced $\mu$-$\tau$ symmetry breaking effect,
no matter which neutrino mass ordering is considered; (b) in the SM
the running direction of $\theta^{}_{23}$ from $\Lambda^{}_{\mu\tau}$
down to $\Lambda^{}_{\rm EW}$ seems to be ``wrong" if one takes today's
best-fit result $\theta^{}_{23} > 45^\circ$ seriously in the normal mass
ordering case; and (c) the SM itself may suffer from the vacuum-stability
problem as the energy scale is above $10^{10}$ GeV \cite{XZZ}.
When the two-Higgs-doublet models (2HDMs) are concerned
\cite{Branco:2011iw}, one may make a similar analysis
to reveal the correlation among the neutrino mass ordering, 
the octant of $\theta_{23}^{}$ and the quadrant of $\delta$ via the
RGE-induced breaking effects. The deviations of $\theta_{23}^{}$
and $\delta$ from their values in the $\mu$-$\tau$ reflection symmetry
limit can be quite different from those in the MSSM case,
depending largely on which 2HDM scenario is taken into account.
Ref. \cite{Zhu} has provided an explicit example of this kind 
in the type-II 2HDM scenario.

On the other hand, we admit that the best-fit values of $\theta^{}_{23}$
and $\delta$ will unavoidably ``fluctuate" in the coming years when more
accurate experimental data are accumulated and incorporated into the
global analysis framework. That is why we have numerically explored the complete
parameter space to illustrate the tolerable ranges of $m^{}_1$ and
$\tan\beta$ which allow us to correlate the normal neutrino mass ordering
with the higher octant of $\theta^{}_{23}$ and the third quadrant of $\delta$.
If the inverted neutrino mass ordering, the lower octant of $\theta^{}_{23}$
and (or) another quadrant of $\delta$ turned out to be favored by the future
precision measurements, it would be straightforward to consider a different
correlation scenario for them either within or beyond the MSSM. In the same
spirit one may study other interesting flavor symmetries and their RGE-induced
breaking, in order to effectively link model building at
high-energy scales to neutrino oscillation experiments at low energies.

\vspace{0.5cm}

{\it We would like to thank Newton Nath, Zhen-hua Zhao, Shun Zhou and Ye-Ling Zhou
for some useful discussions.
This research work was supported in part by the National Natural Science
Foundation of China under grant No. 11775231 and grant No. 11775232.}

\end{document}